# Chapter 3

# Insertion Magnets

*G. Ambrosio[1], M. Anerella[2], R. Bossert[1], D. Cheng[3], G. Chlachidze[1], D. Dietderich[3], D Duarte Ramos[4], P. Fabbricatore[5], S. Farinon[5], H. Felice[3], P. Ferracin[4], P. Fessia[4], J. Garcia Matos[6], A. Ghosh[2], P. Hagen[4], S. Izquierdo Bermudez[4], M. Juchno[4], S. Krave[1], M. Marchevsky[3], T. Nakamoto[7], T. Ogitsu[7], J.C. Perez[4], H. Prin[4], J.M. Rifflet[8], G.L. Sabbi[3], K. Sasaki[7], J. Schmalzle[2], M. Segreti[8], M. Sugano[7], E. Todesco[4]\*, F. Toral[6], G. Volpini[9], P. Wanderer[2], X. Wang[3], R.V. Weelderen[4], Q. Xu[7] and M. Yu[1]*

[1]FNAL, Fermi National Accelerator Laboratory, Batavia, USA  
[2]BNL, Brookhaven National Laboratory, Upton, USA  
[3]LBNL, Lawrence Berkeley National Laboratory, Berkeley, USA  
[4]CERN, Accelerator & Technology Sector, Geneva, Switzerland  
[5]INFN, Sezione di Genova, Genova, Italy  
[6]CIEMAT, Madrid, Spain  
[7]KEK, Tsukuba, Japan  
[8]CEA/SACLAY, DSM/Irfu/SACM, Gif-sur-Yvette, France  
[9]INFN-LASA, Milan, Italy

## 3    Insertion magnets

### 3.1    Overview

The layout of the HL-LHC insertion magnets is shown in Figure 3-1 and compared to those of the LHC in Figure 3-2 . The main technical choices can be summarized as follows [1, 2].

- Maintain the distance from the first magnet to the collision point at ~23 m. This allows preservation of the most critical interfaces with the detectors.

- Increase the triplet coil aperture from 70 mm to 150 mm to allow a smaller $\beta^*$. $Nb_3Sn$ technology has been selected for the quadrupoles [3], allowing an increase in the aperture while keeping the magnet length at acceptable values. The choice of a large coil width (about 36 mm, arranged in two layers of 18 mm wide cable) aims at reaching maximum performance in terms of gradient [1, 4]. At the same time, the operational current is set at 80% of the load line, which is a good compromise between risk and performance [5, 6]. In these conditions, $Nb_3Sn$ can generate an operational field of 140 T/m [7], corresponding to a triplet length of ~35 m, compared to ~25 m in the LHC with an operational gradient of 200 T/m.

- To recover the 10 m of additional space allocated to the triplet, and gain further space for inserting the crab cavities (see Chapters 2 and 4), three steps are taken.

    o   Increase the strength of the separation/recombination dipoles from 26 T·m to 35 T·m, thus reducing the D1–D2 distance to 70 m and recovering 15 m.

    o   Replacing the 20 m long normal conducting magnet D1 operating at 1.28 T with a superconducting 6.27 m long magnet, operating at 5.5 T [8], thus recovering ~14 m.

    o   The power feed for the triplet and D1 is not made through a module placed between D1 and the triplet as in the LHC (indicated by DFB in Figure 3-2), but through a service module on the D2 side

---

[*] Corresponding author: Ezio.Todesco@cern.ch

of D1 (not shown in Figure 3-1). This allows the shifting of D2 towards the IP by a few metres, at the price of having the power feeding the triplet and corrector magnets through D1.

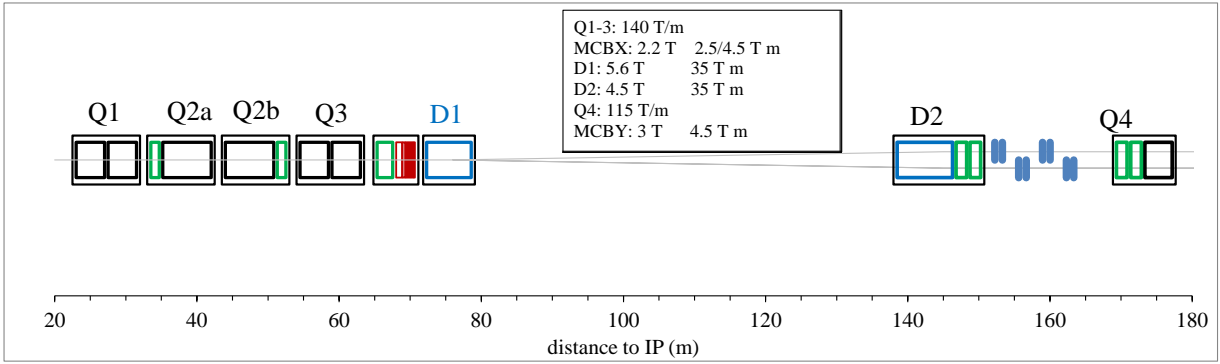

Figure 3-1: Conceptual layout of the IR region of HL-LHC. Thick boxes are magnets, thin boxes are cryostats.

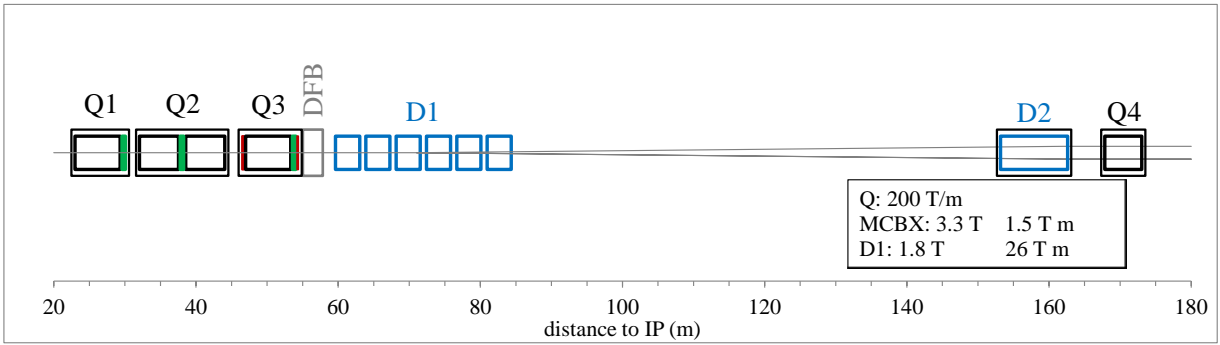

Figure 3-2: Schematic layout of the current IR region of the LHC. Thick boxes are magnets, thin boxes are cryostats.

- Q4 is also shifted by ~5 m away from the interaction point (IP)

- The apertures of the magnets between the triplet and the dispersion suppressor have to be increased: D1 from 60 mm to 150 mm, D2 from 80 mm to 105 mm, Q4 from 70 mm to 90 mm, and Q5 from 56 mm to 70 mm. For all these magnets, Nb-Ti technology has been chosen [1, 2], since the potential performance improvement given by $Nb_3Sn$ is not sufficient to justify the additional cost and complexity. Since the size of the magnet is limited by the cryostat, and the aperture is enlarged, for D1, D2, and Q4 we selected a small coil width (15 mm) to have enough space for an appropriate iron yoke. We opted to reuse the LHC cable in a single-layer configuration to reduce the risks (the cable properties are well known), to ease the schedule (lengths are already available), and simplify protection (quench heaters can be replaced by a dump resistor), at the price of a larger operational current.

- All magnets operate at 1.9 K to have the maximum superconductor performance. This is an important change with respect to the LHC, where the D2, Q4, Q5, and Q6 operational temperature is 4.5 K.

- Three orbit correctors are required in the triplet. The strength is increased from 1.5 T·m (LHC value) to 2.5 T·m for the correctors close to Q2a/b, and to 4.5 T·m for the corrector close to Q3. The position is the same as in the LHC layout, with the exception of the corrector between Q2a and Q2b, which is moved to between Q2b and Q3.

- A skew quadrupole is used to correct the triplet tilt, as in the LHC. Non-linear correctors of the order 3, 4, 5, and 6 are required, both normal and skew. With respect to the LHC layout, normal and skew decapole correctors and a skew dodecapole corrector are added. Experience with LHC operation and field quality of the triplet short models will confirm whether these correctors are needed.

- With a nominal luminosity five times larger than the nominal design goal of the LHC, a newly designed absorber, using thick tungsten (W) shielding attached to the outer surface of the beam screen (Figure 3-3) is foreseen to reduce the effect of collision debris. The tungsten shielding will limit the radiation damage over the HL-LHC accumulated luminosity of 3000 fb$^{-1}$ to a maximum of 30 MGy and the peak energy deposition to a maximum of 4 mW/cm$^3$ [9]. These values are similar to the expected heat load and radiation doses for the nominal LHC [10]. The cryogenic system from the triplet to D1 has to absorb 1.2 kW steady-state at a nominal luminosity of $5 \times 10^{34}$ cm$^{-2}$ s$^{-1}$. Half of this is intercepted by the cold mass at 1.9 K and half by the beam screen, where heat is removed at 40–60 K. Note that the system has to be able to remove a 50% larger load, corresponding to the ultimate peak luminosity of $7.5 \times 10^{34}$ cm$^{-2}$ s$^{-1}$.

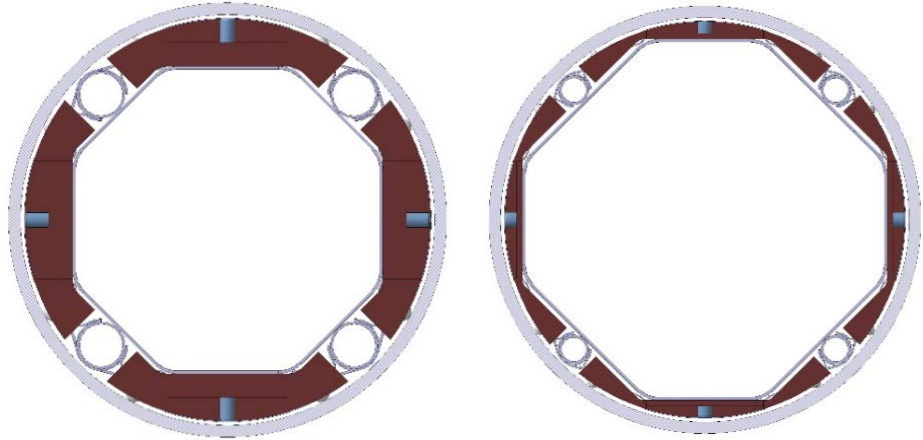

Figure 3-3: Beam screen (grey) with tungsten shielding (dark brown) and cooling tubes in Q1 (lefthand side) and in Q2-D1 (righthand side).

The main parameters of the magnets are listed in Table 3-1.

Table 3-1: Overview of the main parameters of the insertion magnets – see the text for definitions of footnotes

|  | Units | Triplet Q1/Q3 MQXFA | Triplet Q2a/b MQXFB | Short orbit corrector MCBXFB | Long orbit corrector MCBXFA | Separation dipole D1 MBXF | Recomb. Dipole D2 MBRD | Orbit corrector MCBRD | Large ap. 2-in-1 Q4 MQYY |
|---|---|---|---|---|---|---|---|---|---|
| Aperture[1] | [mm] | 150 | 150 | 150 | 150 | 150 | 105 | 100 | 90 |
| Field | [T] |  |  | 2.10 | 2.10 | 5.58 | 4.50 | 3.00 |  |
| Gradient | [T/m] | 140 | 140 |  |  |  |  |  | 115 |
| Magnetic length[2] | [m] | 4.00 | 6.80 | 1.20 | 2.20 | 6.27 | 7.78 | 1.5 | 3.83 |
| Int field | [T m] |  |  | 2.5 | 4.5 | 35.0 | 35.0 | 4.5 |  |
| Int gradient | [T] | 560 | 952 |  |  |  |  |  | 440 |
| Number of apertures |  | 1 | 1 | 1 | 1 | 1 | 2 | 2 | 2 |
| Distance between apertures[1] | [mm] |  |  |  |  |  | 188 | 188 | 194 |
| Number of circuits |  | 4 | 4 | 16 | 8 | 4 | 4 | 32 | 8 |
| Units needed |  | 16 | 8 | 8 | 4 | 4 | 4 | 16 | 4 |
| Spares |  | 4 | 2 | 2 | 2 | 2 | 2 | 2 | 2 |
| Cable data | | | | | | | | | |
| Material |  | Nb$_3$Sn | Nb$_3$Sn | Nb-Ti | Nb-Ti | Nb-Ti | Nb-Ti | Nb-Ti | Nb-Ti |
| Cable thick. in.[3] | [mm] | 1.438 | 1.438 | 0.819 | 0.819 | 1.362 | 1.362 | 0.819 | 1.362 |
| Cable thick. ou.[3] | [mm] | 1.612 | 1.612 | 0.871 | 0.871 | 1.598 | 1.598 | 0.871 | 1.598 |
| Cable width[3] | [mm] | 18.15 | 18.15 | 4.37 | 4.37 | 15.10 | 15.10 | 4.37 | 15.10 |
| Ins. thick radial[4] | [mm] | 0.150 | 0.150 | 0.105 | 0.105 | 0.155 | 0.150 | 0.105 | 0.160 |
| Ins. thick azimuth.[4] | [mm] | 0.150 | 0.150 | 0.105 | 0.105 | 0.135 | 0.130 | 0.105 | 0.145 |
| No. strands |  | 40 | 40 | 18 | 18 | 36 | 36 | 18 | 36 |
| Strand diameter[3] | [mm] | 0.850 | 0.850 | 0.480 | 0.480 | 0.825 | 0.825 | 0.480 | 0.480 |
| Cu/NonCu |  | 1.20 | 1.20 | 1.75 | 1.75 | 1.95 | 1.95 | 1.75 | 1.75 |
| Filling factor[5] |  | 0.31 | 0.31 | 0.25 | 0.25 | 0.24 | 0.24 | 0.25 | 0.25 |

| | | Triplet Q1/Q3 MQXFA | Triplet Q2a/b MQXFB | Short orbit corrector MCBXFB | Long orbit corrector MCBXFA | Separation dipole D1 MBXF | Recomb. Dipole D2 MBRD | Orbit corrector MCBRD | Large ap. 2-in-1 Q4 MQYY |
|---|---|---|---|---|---|---|---|---|---|
| **Coil design** | | | | | | | | | |
| N. layers | | 2 | 2 | 1+1 | 1+1 | 1 | 1 | 1 | 1 |
| N. turns/pole | | 50 | 50 | 74/88 | 74/88 | 44 | 31 | 52 | 14 |
| Cable length/pole | [m] | 430 | 710 | 220/240 | 380/430 | 600 | 500 | 310 | 110 |
| | Units | Triplet Q1/Q3 MQXFA | Triplet Q2a/b MQXFB | Short orbit corrector MCBXFB | Long orbit corrector MCBXFA | Separation dipole D1 MBXF | Recomb. Dipole D2 MBRD | Orbit corrector MCBRD | Large ap. 2-in-1 Q4 MQYY |
| **Operational parameters** | | | | | | | | | |
| Peak field[6] | [T] | 12.1 | 12.1 | 4.3 | 4.3 | 6.6 | 5.1 | 3.8 | 6.0 |
| Temperature | [K] | 1.9 | 1.9 | 1.9 | 1.9 | 1.9 | 1.9 | 1.9 | 1.9 |
| Current | [kA] | 17.46 | 17.46 | 2.42/2.14 | 2.42/2.14 | 11.80 | 12.00 | 3.20 | 15.65 |
| j overall[7] | [A/mm$^2$] | 486 | 486 | 331/290 | 331/290 | 438 | 448 | 662 | 573 |
| Loadline fraction[8] | [adim] | 0.80 | 0.80 | 0.60 | 0.60 | 0.75 | 0.65 | 0.5 | 0.80 |
| Temperature margin | [K] | 4.2 | 4.2 | 3.5 | 3.5 | 2.5 | 3.0 | | 2.0 |
| Stored energy/m | [MJ/m] | 1.32 | 1.32 | 0.100 | 0.100 | 0.342 | 0.284 | 0.174 | 0.190 |
| Inductance/m | [mH/m] | 8.22 | 8.22 | 15.2 / 24.2 | 15.2 / 24.2 | 4.01 | 3.51 | | 0.77 |
| Stored energy[9] | MJ | 5.28 | 8.98 | 0.122 | 0.223 | 2.15 | 2.21 | | 0.73 |
| **Mechanical structure** | | | | | | | | | |
| Forces x | [MN/m] | 2.47 | 2.47 | 0.322 | 0.322 | 1.53 | 0.64 | | 0.33 |
| Forces y | [MN/m] | -3.91 | -3.91 | 0.402 | 0.402 | -0.64 | -0.40 | | -0.47 |
| F$_{mag}$ stress[10] | [MPa] | 130 | 130 | 135 | 135 | 100 | 50 | | 35 |
| **Protection** | | | | | | | | | |
| Circuit inductance[11] | [mH] | 132 | 112 | 18.3/29.1 | 33.5/53.2 | 25 | 27 | | 2.9 |
| Coil energy density[12] | [J/mm$^3$] | 0.092 | 0.092 | 0.026/0.03 | 0.026/0.03 | 0.072 | 0.043 | | 0.062 |
| Dump resistor[13] | [mW] | 50 | 50 | 300 | 300 | 70 | 70 | 250 | 50 |
| Heater circuits[14] | | 12 | 12 | 0 | 0 | 4 or 0 | 4 or 0 | 0 | 0 |
| **Dose and heat load given by collision debris** | | | | | | | | | |
| Coil peak power[15] | [mW/cm$^3$] | 1.1/2.5 | 2.0/2.5 | 1.1/2.5 | 1.1 | 1.3 | 1.8 | 0.6 | 0.3 |
| Heat load cold mass[16] | [W] | 110/160 | 105/130 | | 70 | 100 | 50 | 10 | 6 |
| Heat load beam screen[16] | [W] | 150/60 | 50/70 | | 45 | 50 | <5 | <1 | <1 |
| Peak dose[17] | [MGy] | 11/33 | 22/32 | 28 | 44 | 15 | 36 | 20 | 8 |

[1] Aperture is the coil inner diameter at room temperature, excluding ground insulation, cold bore, and beam screen; distance between apertures is given at 1.9 K.
[2] Magnetic length is given at 1.9 K.
[3] Strand/cable dimensions are given at room temperature, in the case of Nb$_3$Sn before reaction.
[4] Insulation dimensions are given at room temperature.
[5] Filling factor is defined as the fraction of superconductor in the insulated cable.
[6] Peak field in the coil is given including the contribution of the strand where the peak is located (what is usually called self-field correction).
[7] Overall current density is the average over the whole cross-section of the insulated cable at 1.9 K (i.e. including voids or impregnation and insulation, but not copper wedges).
[8] Load line fraction is the ratio between the operational current and the short sample current on the load line.
[9] Stored energy is given for the whole magnet: in the case of independently powered apertures or nested magnets, stored energy is given for both circuits powered with maximum nominal current.
[10] Stress is an estimate given by the accumulation of the azimuthal Lorentz forces at nominal current divided by the coil radial width – the impact of the structure, preload, and bending is not considered.
[11] Circuit inductance is the differential inductance of the circuit at maximum nominal current.
[12] Energy density is given over the coil volume at 1.9 K, including insulation but not coil parts such as copper wedges and pole pieces.
[13] Dump resistor is estimated using a maximum voltage target of ~750 V.
[14] The number of heater circuits is the number of independent circuits for each magnet.
[15] Peak power is the maximum deposited power in the coil for a nominal luminosity of $5 \times 10^{34}$ cm$^{-2}$ s$^{-1}$ (details on the binning and simulations are given in Chapter 10).
[16] Heat load is given for nominal luminosity $5 \times 10^{34}$ cm$^{-2}$ s$^{-1}$, separated between cold mass (this portion is absorbed at 1.9 K) and beam screen (where it is absorbed at 40–60 K); values for the short orbit correctors and for Q2a and Q2b are given together, in the Q2a Q2b column; the load on the interconnections is given in the text. For ultimate performance these values are 50% large.
[17] Peak dose is the same as the maximum dose in the magnet (for details of binning and simulations see Chapter 10).

### 3.2 Triplet

**Function and operational modes**: The triplet ramps with the energy of the LHC, with a nominal gradient of 9 T/m at 450 GeV, and 140 T/m at 7 TeV. During squeeze, its gradient is constant or decreases by not more than 10%. Q1 and Q3 will be operated in series, as will Q2a and Q2b. A 2 kA powering trim is acting on Q3: this is needed for special beam measurements requiring different power between Q1 and Q3. The quadrupoles of the triplet are developed by LARP and CERN.

**Cable**: A $Nb_3Sn$ cable based on 40 strands, with 0.85 mm diameter, has been chosen [7]. The main specifications are (i) the minimum non-copper critical current density of 1400 A/mm$^2$ at 15 T and 4.2 K, (ii) a strand RRR larger than 150, and (iii) a Cu/no Cu ratio of 1.2 (54.5% of copper in the strand). The cable has a S2 glass braided insulation, whose thickness is 150 μm at 5 MPa before reaction. The cable contains a 12-mm-wide, 25-μm-thick stainless steel core to control and reduce the dynamic effects.

**Coil, current density, and margin**: With two layers one can reach the operational gradient of 140 T/m at 0.80 of the short sample limit on the load line (i.e. 20% of load line margin). Each layer has a copper wedge to tune field quality.

**Lengths and transverse size**: The triplet is made of Q1 and Q3, each unit requiring magnetic length of 8 m; plus two units Q2a and Q2b, each one with a 6.8 m long magnetic length. The LARP collaboration, in charge of Q1 and Q3 construction, has proposed splitting both Q1 and Q3 into two parts, each resulting in 4 m long magnets separated by 0.5 m. The Q1, Q2, and Q3 cross-sections are identical, and make use of the same design, technologies, and components. The magnet cross-section has a 630 mm diameter, i.e. 60 mm more than the LHC dipoles, including the stainless steel vessel.

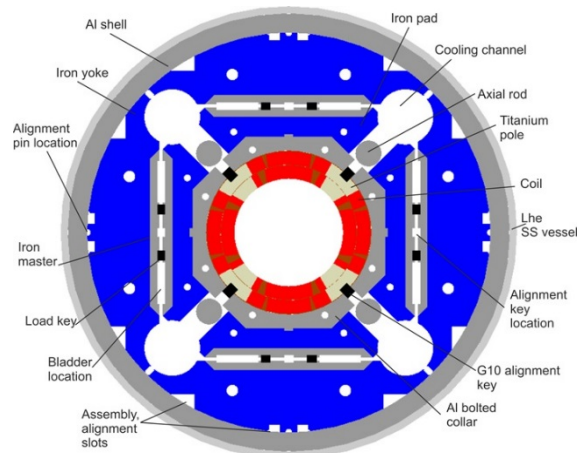

Figure 3-4: Sketch of triplet quadrupole magnet cross-section

**Mechanical structure**: The quadrupole makes use of a shell-based structure developed at LBNL and within the LARP collaboration [11]. The structure, scaled up to a length of 3.4 m, was demonstrated in the LARP LR and LQ quadrupole models [12], and features to assure alignment in operational conditions have been included in the 120 mm aperture quadrupole HQ [13]. The QXF cross-section is a scaling up of the HQ design. Coils are mainly prestressed by the Al shell during cool-down, acting as the structure to contain the Lorentz forces during powering up. The level of stress is fine-tuned during the loading of the coil, which is done at room temperature using water-pressurized bladders and interference keys. Typically one has ~70 MPa of azimuthal coil compression at room temperature, which becomes ~150 MPa at 1.9 K thanks to the interplay of the thermal contractions of the different components. The structure keeps the coil under compression up to the ultimate current, corresponding to 150 T/m (7% above nominal).

**Protection**: The energy density in the coil is ~0.1 mJ/mm$^3$, about twice that of the Nb-Ti LHC magnets [14]. This makes quench protection particularly challenging. Since the circuit inductance is of the order of 100 mH, only a small fraction of the energy can be extracted on a dump resistor. Therefore we have to rely on quench heaters on the outer layer of the coil, i.e. 25 µm stainless steel strips with a 50 µm polyamide layer to ensure proper insulation. The heaters will have heating stations of 40 mm length, separated by 120 mm sections with lower resistance due to a 10 µm copper cladding (see Figure 3-5(a). Since the width of the heating stations is 20 mm, two independently powered strips will cover the two blocks of the outer layer; a 6.8 m long magnet will have ~40 heating stations. The typical time needed to quench the coil is of the order of 15–20 ms following heater firing [15]. Assuming 5 ms for detection time, a validation window of 10 ms and a few ms for switch opening, this brings the hotspot temperature to ~350 K [16]. To reduce this value and to ensure some redundancy, we plan to also have heaters on the inner layer, and/or use the CLIQ system [17], recently developed at CERN, based on coil heating induced by fast current discharge in the magnet. For the inner layer, heating stations are also needed, with a more complex geometry since ~50% of the surface must be left free for heat removal. A single strip slaloms between the two blocks, with 25 mm long heating stations and 40 mm long cladding (see Figure 3-5(b). The CLIQ system has the interesting feature of acting rapidly on the inner layer, and is therefore complementary to the outer layer quench heaters. It requires an additional lead to bring the current between coil poles: the viability of this protection system in the case of more complex circuits (two magnets in series, with an additional trim) is to be carefully studied.

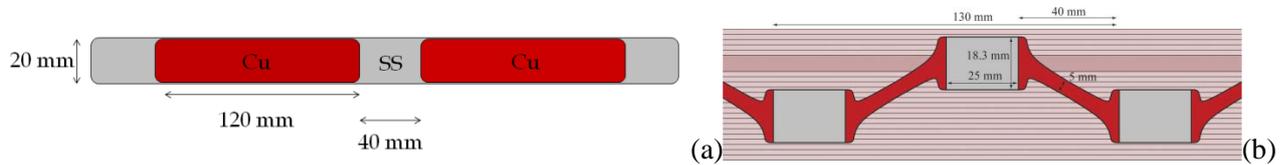

Figure 3-5: (a) Quench heaters for the outer layer; (b) a design for the inner layer. Stainless steel (SS) in grey and copper cladding in red.

**Field quality**: Allowed field harmonics ($b_6$, $b_{10}$) are optimized at high field, and will be below one unit in absolute value. Contributions from the coil ends have to be taken into account and optimized and/or compensated through the straight part [18]. Fine tuning of $b_6$ can be done in the short model phase through small changes to the coil cross-section. Random components are estimated for a 25 µm random error in the block positioning for non-allowed, and 100 µm for allowed (see Table 3-2); most critical components are low-order harmonics ($b_3$, $a_3$, $b_4$, $a_4$). To minimize these components we opted for a strategy similar to that used in the RHIC magnets [19], with magnetic shims to be inserted in the bladder location [20]. This allows correcting (i) ±5 units of $b_3$; (ii) ±5 units of $a_3$; (iii) ±3 units of $b_4$; (iv) ±1 units of $a_4$, for a maximum of two harmonics at the same time, through an asymmetric magnetic shimming.

Table 3-2: Expected systematic harmonics and random components in the triplet

| Expected systematic harmonics | | | | | | Random components | | |
|---|---|---|---|---|---|---|---|---|
| | Geometric | Saturation | Persistent | Injection | High field | Order | Normal | Skew |
| $b_6$ | 4.20 | −3.80 | −20.00 | −15.80 | 0.40 | 3 | 0.82 | 0.80 |
| $b_{10}$ | −0.37 | −0.02 | 4.00 | 3.63 | −0.39 | 4 | 0.57 | 0.65 |
| $b_{14}$ | −0.60 | −0.07 | 0.00 | −0.60 | −0.67 | 5 | 0.42 | 0.43 |
| | | | | | | 6 | 1.10 | 0.31 |
| | | | | | | 7 | 0.19 | 0.19 |
| | | | | | | 8 | 0.13 | 0.11 |
| | | | | | | 9 | 0.07 | 0.08 |

**Cooling**: The magnet is in a static bath of pressurized HeII, with a welded stainless-steel shell placed outside the Al structure acting as a helium vessel. Cooling is ensured via two heat exchangers of 68 mm inner diameter, in which saturated HeII circulates, housed in the 77 mm diameter holes of the iron located in the upper part, see Figure 3-4 [21]. This circuit cools the triplet and the short orbit correctors, with the separation dipole and corrector package on a different circuit (see Section 3.11 for more details). With this design, one can comfortably remove ~800 W, and ultimately ~1000 W, of heat load on the triplet, i.e. 500 W on the cold mass given by debris (see Table 3-1), plus a 100 W budget for other loads (among them the 25 W load on interconnections), with a 50% margin. For the Nb-Ti coils in the LHC, the peak heat deposition target was set at 4 mW/cm$^3$; this has a factor of 3 safety on 12 mW/cm$^3$, which was considered to be the hard limit. Later experience showed that the hard limit is a factor of two larger. In the HL-LHC, thanks to the tungsten shielding, we are always below the 4 mW/cm$^3$ target. Considering that the Nb$_3$Sn superconductor is expected to have a three times larger margin, peak power in the HL-LHC will not be critical.

The heat loads from the coils and the beam-pipe area can only be evacuated to the two heat exchangers by means of pressurized HeII. To this end the cold mass design incorporates the required helium passages: 1.5 mm annular spacing between cold bore and inner coil-block, and free passage through the coil pole and subsequent G10 alignment key. The free passage needed through the coil pole and G10 alignment key in the transverse direction is estimated to be at least 4% (equivalent to 8 mm diameter holes repeated every 40 mm).

**Cryostat**: Independent cryostats are used for Q1, Q2a, Q2b, and Q3. The Q1 and Q3 cryostats contain two 4 m long magnets. The Q2a and Q2b cryostats contain the 6.8 m long magnet plus the orbit correctors described below. The cryostat size should be able to accommodate the cold mass, the shielding, and the cooling pipes. First estimates show that the standard size of 980 mm (including flanges) is a very tight fit for all of these components. Two options are being considered [22]: a 1050 mm diameter cryostat, and an elliptic cryostat (see Figure 3-6).

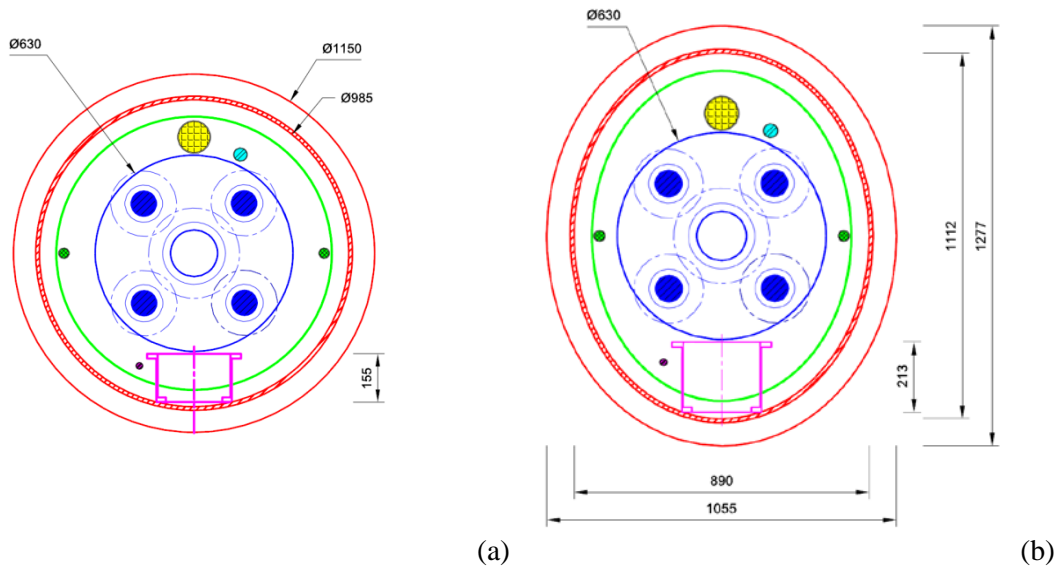

(a)          (b)

Figure 3-6: Possible cross-sections of the cryostat: (a) larger diameter and (b) elliptical solution

### 3.3 Orbit correctors

**Function and operational modes**: The orbit correctors are single-aperture magnets. Two versions are required, providing 2.5 T·m and 4.5 T·m integrated fields. To have a more compact layout, a nested design has been adopted, with the horizontal and vertical dipole coil in the same cross-section (see Figure 3-7). The field in each plane has been limited to 2.1 T, giving a maximum combined field of 3 T. Powering will be allowed in a square in the horizontal/vertical plane, with both positive and negative currents. These magnets are used to create the crossing angle and partially to correct quadrupole misalignment. Orbit correctors are being developed by CIEMAT [23].

**Cable**: The 4.5 mm wide Nb-Ti cable developed for the S-LHC corrector [24] has been adopted. This is based on a 0.48 mm diameter strand, arranged in a Rutherford cable with 18 strands.

**Coil, current density and margin**: With one layer per dipole direction, it can reach the operational field of 2.1 T simultaneously in both planes at 60% of the load line.

**Lengths and transverse size**: The magnetic length is 1.2 m for the short version (MCBXB) and 2.2 m for the long one (MCBXA). The magnet cross-section has a 570 mm diameter, including the stainless steel vessel, i.e. as in the LHC dipoles and D1. Cooling channels are shared with the quadrupoles for MCBXB, so the iron has 77 mm diameter holes at 45°. MCBXA will share cooling channels with D1, thus the iron has 60 mm diameter holes as in D1 (i.e. at 90° and 270º).

**Mechanical structure**: The magnet makes use of an iron shell support and collars. An aluminium shell is used to increase the rigidity of the assembly, applying a pre-stress through the iron that will increase during cool-down. Approximately 20 mm thick collars are used for keeping the inner coil, and the same for the outer coil. Due to the nested option, a complex collaring based on two steps (first the inner, then the outer) is needed. The inner collars are closed with two round pins; the outer ones will be kept in place by four prismatic keys. A difficulty is that when both horizontal and vertical coils are powered, Lorentz forces push the inner coil towards the centre of the aperture: this requires a structure between the inner coil and the cold bore to prevent movement. A titanium tube has been proposed as a solution, given that its contraction coefficient is lower than those of the other materials. All of these issues are being addressed at the time of writing.

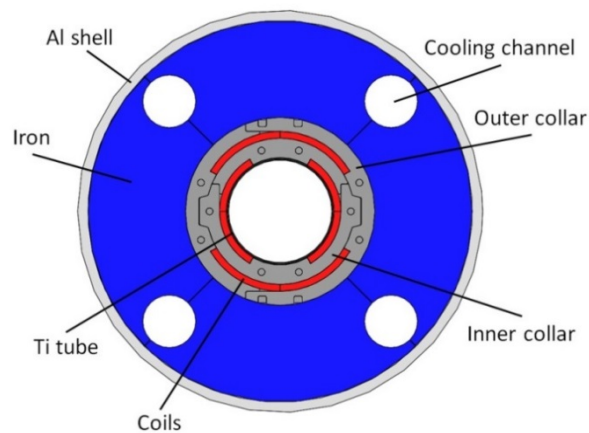

Figure 3-7: Sketch of orbit corrector cross-section (short version MCBXB, in the cold mass of the triplet magnets Q2a and Q2b).

## 3.4  High-order correctors

**Function, design and operational modes**: The high-order correctors (skew quadrupole, normal and skew sextupole, octupole, decapole, and dodecapole) are specified on the expected field quality and alignment errors (see Chapter 2), with a safety factor of 2 on the quadrupole, sextupole, and octupole, and 1.5 on the decapole and dodecapole. The magnets will operate with settings based on the measured field errors of the triplet and separation dipole. To ease operation a non-nested layout (see Figure 3-8) has been adopted, with a superferric technology, already developed for the S-LHC [25]. Nb-Ti racetrack coils provide the ampere-turns, with iron giving the required field shape. The aperture is 150 mm, as for the triplet and D1 (see Figure 3-9). The high-order correctors are being developed by INFN-Milano.

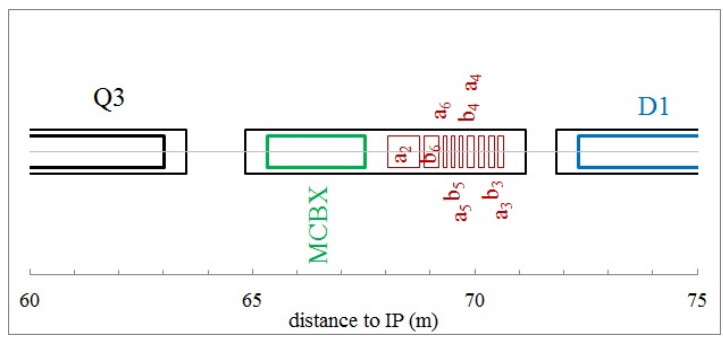

Figure 3-8: Layout of the corrector region

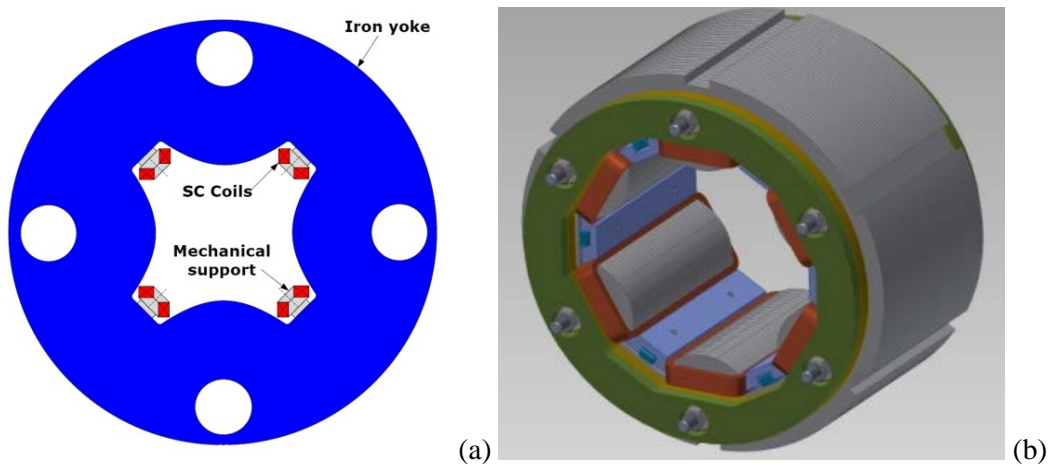

Figure 3-9: (a) Cross-section of the skew quadrupole; (b) 3D view of a sextupole

Cable: the cable is a single Nb-Ti strand, of 0.7 mm diameter for the quadrupole and of 0.5 mm diameter for the higher orders. Insulation is made with a 0.07 mm thick S2 glass. Further ground insulation is added on the external side of the coil.

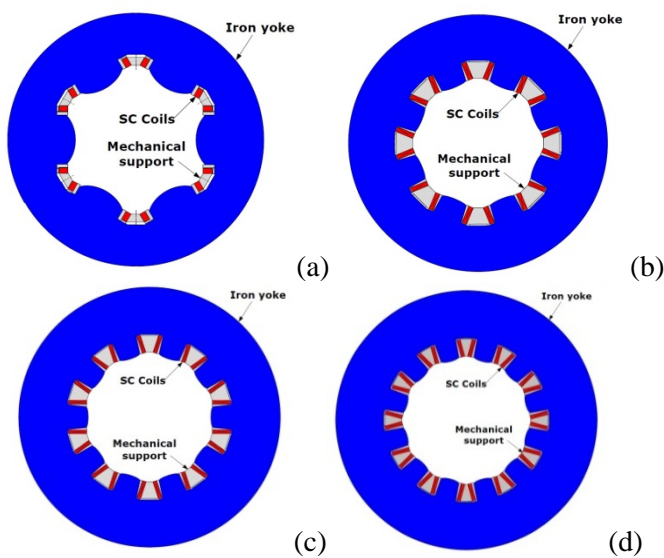

Figure 3-10: Sketch of nonlinear corrector cross-sections of (a) normal sextupole; (b) octupole; (c) decapole; (d) dodecapole correctors.

**Coil, current density, and margin**: We chose to operate at 40% on the load line. The optimization [26] provides a current density of the order of 300 A/mm$^2$, with peak fields on the coil in the range 2.0–2.3 T for the nonlinear correctors and 3.0 T for the skew quadrupole (see Table 3-3). Coils are impregnated with CTD-101.

**Lengths and transverse size**: The magnet coil lengths are ~0.1 m for the sextupole, octupole, decapole, and skew dodecapole (see Table 3-3). The normal dodecapole and the skew sextupole require greater strengths, giving a coil length of ~0.45 m and ~0.84 m, respectively. The skew quadrupole needs a 460 mm diameter iron yoke that has to include the cooling holes for the D1 heat exchanger and orbit corrector. The nonlinear correctors can have an iron yoke diameter of 320 mm, which fits inside the cooling holes. Spacers are required to match the transverse size of the correctors to the same value of the MCBXA orbit correctors, and to maintain alignment within the cold mass. Heat exchangers will go through these spacers to cool the whole cold mass.

**Mechanical structure**: the mechanical support for the correctors coils are under study. The forces are of the order of 60 kN/m for the quadrupole and 10–30 kN/m for the other magnets.

The main parameters of the correctors are given in Table 3-3.

Table 3-3: Overview of the main parameters of the triplet corrector magnets

|  | Units | MCQSX | MCSX/MCSSX | MCOX/MCOSX | MCDX/MCDSX | MCTX | MCTSX |
|---|---|---|---|---|---|---|---|
| Order |  | 2 | 3 | 4 | 5 | 6 | 6 |
| Aperture | [mm] | 150 | 150 | 150 | 150 | 150 | 150 |
| Integrated strength[1] | [T m] | 1 | 0.06 | 0.04 | 0.03 | 0.086 | 0.017 |
| Coil length[2] | [mm] | 841 | 123 | 99 | 107 | 449 | 102 |
| Gradient | [T/m$^{n-1}$] | 25 | 11 | 3690 | 50600 | 640000 | 613000 |
| Number of apertures |  | 1 | 1 | 1 | 1 | 1 | 1 |
| Number of circuits |  | 1 | 2 | 2 | 2 | 1 | 1 |
| Units needed |  | 4 | 8 | 8 | 8 | 4 | 4 |
| Spares |  | 2 | 2 | 2 | 2 | 2 | 2 |
| **Cable data** | | | | | | | |
| Strand diameter | [mm] | 0.700 | 0.500 | 0.500 | 0.500 | 0.500 | 0.500 |
| Insulation thickness | [mm] | 0.070 | 0.070 | 0.070 | 0.070 | 0.070 | 0.070 |
| Cu/No_Cu |  | 2.3 | 2.3 | 2.3 | 2.3 | 2.3 | 2.3 |
| **Coil design** | | | | | | | |
| Material |  | Nb-Ti | Nb-Ti | Nb-Ti | Nb-Ti | Nb-Ti | Nb-Ti |
| N. turns/pole |  | 320 | 214 | 344 | 256 | 154 | 172 |
| Cable length/pole | [m] | 604 | 79 | 88 | 67 | 144 | 42 |
| **Operational parameters** | | | | | | | |
| Coil peak field | [T] | 2.97 | 2.33 | 2.41 | 2.34 | 2.04 | 2.01 |
| Temperature | [K] | 1.9 | 1.9 | 1.9 | 1.9 | 1.9 | 1.9 |
| Current | [A] | 182 | 132 | 120 | 139 | 167 | 157 |
| j overall[3] | [A/mm$^2$] | 303 | 353 | 314 | 360 | 259 | 284 |
| Loadline fraction |  | 0.40 | 0.40 | 0.40 | 0.40 | 0.40 | 0.40 |
| Differential inductance | [mH] | 1247 | 118 | 152 | 107 | 229 | 52 |
| Stored energy | [kJ] | 24.6 | 1.24 | 1.41 | 1.39 | 4.35 | 0.92 |
| **Dose and heat load** | | | | | | | |
| Heat load cold mass | [W] | 70 | | | | | |
| Heat load beam screen | [W] | 45 | | | | | |
| Peak dose | [MGy] | 44 | | | | | |

[1] Integrated strength is defined as the field at the 50 mm reference radius times the magnetic length.
[2] Coil length refers to the physical coil length, and not to magnetic length.
[3] The overall current density includes 0.07 mm thick strand insulation and the coil ground insulation.

## 3.5 Separation dipole D1

**Function and operational modes**: The separation dipole is ramped with the energy of the LHC, and is constant during squeeze. Each magnet is independently powered. The separation dipole is being developed by KEK.

**Cable**: The 15 mm wide Nb-Ti cable used for the outer layer of the main LHC dipole is adopted (see Figure 3-10). The required unit length is smaller than the main LHC dipole outer cable unit length (780 m).

**Coil, current density and margin**: For the initial choice of 70% operational level, the magnet length was slightly longer than the KEK test station [8, 27]. We therefore fixed the operational current at 75% of the load line. This allows reaching an operational field of 5.58 T and a magnetic length of 6.27 m, fitting the vertical test station without significantly increasing the risk related to the lower margin.

**Lengths and transverse size**: The magnet cross-section has a 570 mm diameter, including the stainless steel vessel, i.e. the same as the LHC dipoles. A larger diameter has been excluded to be able to reuse the yoking tooling used for J-PARC at KEK.

**Mechanical structure**: Forces are kept by the iron yoke, with thin spacers between the iron and the coil, as the J-PARC [28], RHIC magnets [29], and LHC Q1/Q3 [30]. Here the prestress is given by the iron laminations, horizontally split, that are locked through keys (see Figure 3-11). A thin stainless-steel collar acts as a spacer between the coil and the iron yoke. An average prestress of 90 MPa is given at room temperature during the so-called 'yoking'. During cool-down the prestress lowers to 70 MPa, which is enough to counteract the Lorentz forces during powering.

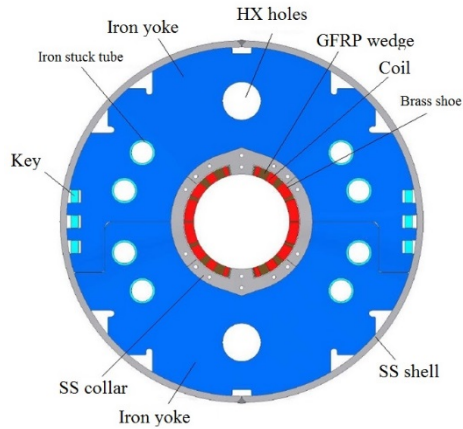

Figure 3-11: Sketch of separation dipole cross-section

**Protection**: The large cable and operational current of 12 kA (see Table 3-1) will probably allow safe operation with energy extraction at a dump and without quench heaters. Simulations show that with a 75 mΩ resistor the hotspot temperature in the adiabatic approximation is 220 K for the nominal current of 12 kA. Quench heaters or CLIQ [17] are still being considered to add a margin and redundancy.

**Field quality**: The main issue here is the saturation component [8], which is optimized via the iron shaping (see Table 3-4). Following the analysis given in Refs. [31, 32], the random components are estimated through random positioning of the coil block with different amplitudes for each family of harmonics, namely 40 µm for the allowed $b_{2n+1}$, 30 µm for the even skew $a_{2n}$, 15 µm for the odd skew $a_{2n+1}$, and 10 µm for the even normal multipoles $b_{2n}$.

**Cooling**: The magnet is in a static bath of pressurized HeII, with a stainless steel shell acting as a helium vessel. Cooling is ensured via two heat exchangers, of 49 mm inner diameter, housed in the 60 mm diameter holes through the iron.

**Cryostat**: The cryostat has the same geometry as the triplet cryostat (see Figure 3-6).

Table 3-4: Expected systematic harmonics and random components of D1

| Expected systematic harmonics ||||| | Random components |||
|---|---|---|---|---|---|---|---|---|
| | Geometric | Saturation | Persistent | Injection | High field | Order | Normal | Skew |
| $b_3$ | −1.80 | 0.90 | −14.2 | −16.00 | −0.90 | 2 | 0.20 | 0.68 |
| $b_5$ | 0.50 | −0.50 | −1.00 | −0.50 | 0.00 | 3 | 0.73 | 0.28 |
| $b_7$ | 1.60 | −1.20 | −0.70 | 0.90 | 0.40 | 4 | 0.13 | 0.44 |
| $b_9$ | −0.68 | 0.09 | 0.02 | −0.66 | −0.59 | 5 | 0.36 | 0.15 |
| $b_{11}$ | 0.44 | 0.03 | 0.00 | 0.44 | 0.47 | 6 | 0.06 | 0.18 |
| | | | | | | 7 | 0.16 | 0.06 |
| | | | | | | 8 | 0.03 | 0.06 |
| | | | | | | 9 | 0.06 | 0.02 |
| | | | | | | 10 | 0.01 | 0.03 |
| | | | | | | 11 | 0.02 | 0.01 |

## 3.6 Recombination dipole D2

**Function and operational modes**: The recombination dipole is ramped with the energy of the LHC, and is constant during squeeze. Each magnet is independently powered, and the two apertures are in series. The fields point in the same direction in both apertures; this makes field quality control much more challenging than in the LHC dipole, where the field points in opposite directions. The design of the recombination dipole is being studied by INFN-Genova.

**Cable**: The 15 mm wide Nb-Ti cable used for the outer layer of the main LHC dipole is adopted. The required unit length is not larger than that of the LHC main dipole's outer layer unit length (780 m).

**Coil, current density and margin**: We selected a conservative margin, operating at 65% of the load line with a single layer 15 mm width coil, and an operational field of 4.5 T. In these conditions, the approach used in the current D2, using iron to magnetically decouple the two apertures, leads to large saturation effects. An alternative approach using left–right asymmetric coils was therefore adopted [33] to compensate for the cross-talk between the two apertures (see Figure 3-12). A very careful optimization is needed to find the best solution. After several iterations, a cross-section was found where the left–right asymmetry is only given by the angles of the blocks, but the number of cables per block is the same [34]. This allows for much simpler coil heads.

**Lengths and transverse size**: The magnetic length is 7.78 m. The magnet requires an adequate iron thickness to reduce the fringe field. An elliptical iron yoke is proposed, of 570 mm vertically and 630 mm horizontally.

**Mechanical structure**: The square design of the central aperture in the iron is imposed by field quality optimization, namely the reduction of the field harmonics due to saturation. The accumulation of Lorentz forces corresponds to a pressure in the midplane of about 40 MPa. A self-supporting stainless-steel twin collar, where the iron is only for alignment, has been shown to be viable, with a peak stress during collaring of the order of 100 MPa [35]. An alternative structure with separated stainless steel collars, allowing more flexibility in production, is also being considered.

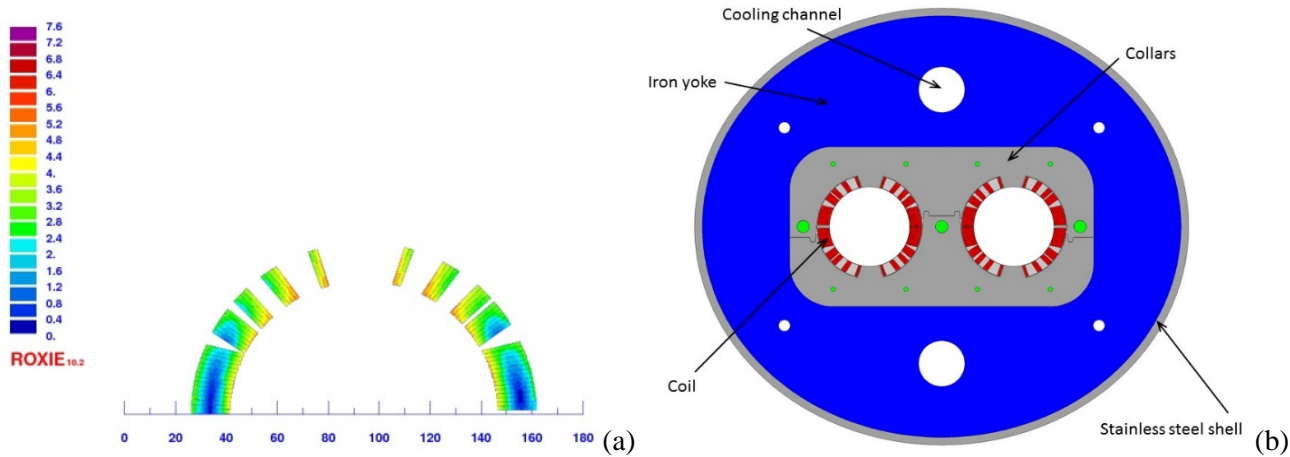

Figure 3-12: Sketch of recombination dipole cross-section. (a) Asymmetric coil; (b) magnet cross-section.

**Protection**: Protection will be probably based on quench heaters; studies are in progress at the time of writing.

**Field quality**: This is the main issue for this magnet. Cross-talk is optimized via the asymmetric cross-section, and the saturation through an iron shaping. Field quality tables are still in progress at the time of writing.

**Cooling**: The magnet is in a static bath of pressurized HeII. Cooling is ensured via two heat exchangers, of 51 mm diameter, housed in a 60 mm diameter hole in the iron, located in the upper part.

### 3.7 D2 and Q4 correctors

**Function and operational modes**: Both D2 and Q4 need orbit correctors for each beam and each plane (horizontal and vertical), with an integrated strength of 4.5 T·m. An aperture of 100 mm is required. In the preliminary layout, a field of 3 T for a magnetic length of 1.5 m has been allocated. These correctors are used to open the crossing angle and to correct the closed orbit, and therefore they should be powered in any configuration.

**Cable**: The 4.5 mm wide Nb-Ti cable used for the single aperture orbit corrector is considered.

**Coil, current density, and margin**: With one layer, one can reach the operational field of 3 T with a comfortable margin (~40%) on the load line.

**Field quality**: The challenge in these magnets is the cross-talk between the apertures. Since the beam distance is 194 mm, and the aperture is 100 mm, little space is left for the iron to decouple the two apertures [36]. No optimization can be made through the iron or the coil cross-section as is the case in D2, since these magnets have to be powered with any combination of currents. The solution is to keep the coil width as small as possible and to have thin collars to maximize the iron thickness. Requirements on field quality are stringent, especially on $b_3$ where the tolerance is of the order of 1.5 units. Given these constraints, the operational field will probably have to be lowered to reduce the cross-talk. Special ferromagnetic laminations are also considered as an option for further decreasing the coupling between the apertures.

**Mechanical structure**: 15 mm thick collars would provide a self-supporting structure, allowing the iron to align the two modules, and to shield the fields.

**Cooling**: The magnets will share cooling with D2 and Q4, so will have heat exchangers in the same position.

## 3.8 Large aperture two-in-one quadrupole Q4

**Function and operational modes**: The Q4 quadrupole (MQYY) is ramped with the energy of the LHC. During the squeeze the gradient is lowered and a minimum operational current of 3% of nominal is required, to allow flexibility for the optics. The two apertures are independently powered, with a possible current unbalance up to 50%. The gradients shall have the same sign in both apertures, resulting in a focusing effect for one beam and defocusing for the other one. A Q4 short model, only, however, of single aperture, is being developed by CEA-Saclay.

**Cable**: The 15 mm wide Nb-Ti cable used for the outer layer of the main LHC dipole is adopted. The required unit length is much smaller than the LHC unit length (780 m), so short lengths of the production that could not be used in the dipoles can be employed, at zero cost, and not affecting the LHC dipole spares. Insulation is based on the enhanced scheme proposed in Ref. [37], allowing a direct path from the cooling bath to the strand.

**Coil, current density and margin**: With one layer, this can reach the operational gradient of 115 T/m at 80% of the load line [38].

**Lengths and transverse size**: The magnetic length is 3.83 m. The magnet cross-section has a 497.5 mm nominal diameter, including the stainless steel vessel, i.e. the same as the LHC main quadrupoles.

**Mechanical structure**: The quadrupole makes use of self-standing collars, the same as the main LHC quadrupoles [39] and MQXC [40] (Figure 3-13). Here, the prestress is given by the 25 mm thick thick collars, which are locked through keys. An average prestress of 80 MPa is given at room temperature after yoking. Thirty percent of prestress is then lost due to cable insulation creep, leading to 55 MPa. During cool-down the prestress lowers to 40 MPa, enough to counteract the Lorentz forces during powering. The iron is used to magnetically separate the two apertures, close the flux lines, and provide alignment. The shells will probably be welded with the quadrupole and the correctors, since an inertia tube would become too long (8 m) for the vertical assembly.

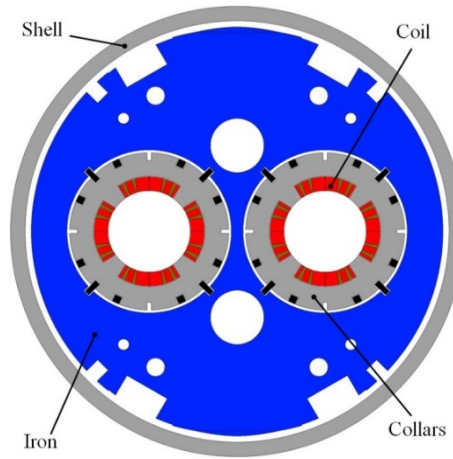

Figure 3-13: Sketch of Q4 cross-section

**Protection**: The large cable and operational current will allow safe operation, with energy extraction in a dump. Simulations show that with a 50 mΩ resistor the hotspot temperature in the adiabatic case is below 200 K.

**Field quality**: The coil has three blocks, allowing $b_6$ and $b_{10}$ to be within one unit and $b_{14}$ to be within two units (see Table 3-5). Shimming on the pole and in the midplane is foreseen to allow the steering of $b_6$ in the model/prototype phase. An estimate of the random component is given in Table 3-5.

**Cooling**: The magnet is in a static bath of HeII, with a stainless steel inertia tube acting as a helium vessel. Cooling is ensured via heat exchangers of 51 mm diameter, housed in the 60 mm diameter holes in the iron, located in the upper part.

**Cryostat**: The cryostat will include the quadrupole plus two orbit correctors (see the previous section), with a total length of about 8 m.

Table 3-5: Expected systematic harmonics and random components

| | Expected systematic harmonics | | | | | | Random components | | |
|---|---|---|---|---|---|---|---|---|---|
| | Geometric | Saturation | Persistent | Injection | High field | | Order | Normal | Skew |
| $b_6$ | −0.45 | 0.40 | −11.00 | −11.45 | −0.05 | | 3 | 1.79 | 1.79 |
| $b_{10}$ | 0.00 | 0.00 | 1.00 | 1.00 | 0.00 | | 4 | 1.16 | 1.16 |
| $b_{14}$ | 1.50 | 0.00 | 0.00 | 1.50 | 1.50 | | 5 | 0.75 | 0.75 |
| | | | | | | | 6 | 1.93 | 0.48 |
| | | | | | | | 7 | 0.31 | 0.31 |
| | | | | | | | 8 | 0.20 | 0.20 |
| | | | | | | | 9 | 0.13 | 0.13 |
| | | | | | | | 10 | 0.34 | 0.08 |

## 3.9  Q5

For the Q5 magnet, the present baseline is to replace it with a 70 mm aperture MQY [41] (at the time of writing Q4 around IP1 and IP5), operating at 1.9 K. With an operational current of 4510 A, it reaches an operational gradient of 200 T/m. As for Q4, apertures are independently powered. There are seven available spares, plus four MQY that will be recovered from IP1 and IP5 in the Q4 position. An alternative option, requiring a larger aperture, is also under consideration. In this case the 90 mm aperture Q4 would be also used as Q5, with a longer length or with two units.

## 3.10  Powering

The baseline of the powering scheme of the triplet-corrector-separation dipole is shown in Figure 3-14. The powering is from the D2-side of D1, allowing a more compact layout with respect to the LHC, where this is done through a distribution feedbox taking a few metres between the triplet and D1. This choice improves performance at the price of having triplet and corrector cables going through (or along) the separation dipole. The second important choice is that magnets are fed by a superconducting link: therefore the transition from superconducting to resistive leads is shifted from the neighbourhood of the beam line. For the triplet, the baseline option is to have Q1 and Q3 in series, and Q2a and Q2b in series, with two 20 kA independent power converters, plus a 2 kA trim on Q3. This configuration provides a good compromise between cost and operation. Other options are under consideration at the time of writing.

- Power in series Q1 and Q2a, and Q2b and Q3: this configuration allows compensation of the power converter ripple between a couple of focusing/defocusing quadrupoles, at the price of a smaller flexibility in the optics.

- All four magnets in series, with a single 20 kA power converter. In this case a 2 kA trim is needed on Q1 and a second trim on Q3. This configuration minimizes the cost of the hardware, but is the more complex from the point of view of operation. It also provides the largest inductance per circuit, increasing the challenges related to the busbar and magnet protection.

- Four independent circuits for Q1, Q2a, Q2b, and Q3, with four 20 kA power converters. This solution provides the best flexibility for optics and operation, at a higher hardware cost. It also provides the lowest inductance per circuit, easing protection of the busbar and of the magnet.

Besides the triplet, one has one 13 kA circuit for D1, and six 3 kA circuits for the orbit correctors, plus the nine correctors rated at 200–300 A. In the baseline case, with 20 kA circuits for the triplet, one has a total of ~70 kA of current to bring in and out of the triplet–D1 area through the link (see Chapter 6). For the busbar the baseline is to place it inside the magnet (as in the LHC main cell), through the iron holes that are not used by the heat exchangers; two other options are also under study:

- busbar inside the cryostat, but outside the cold mass, in a separate line (as the M line in the LHC cell, carrying some correctors busbars);
- busbar outside the cryostat in a separate cryostat, in this case each magnet is fed by a bypass of the busbar cryostat.

The matching section is fed by a second superconducting link, bringing current to the recombination dipole D2, Q4, and the orbit correctors. Here we have one 13 kA circuit for D2, two 20 kA circuits for Q4, and four 3 kA circuit for the orbit correctors, with a total of 65 kA.

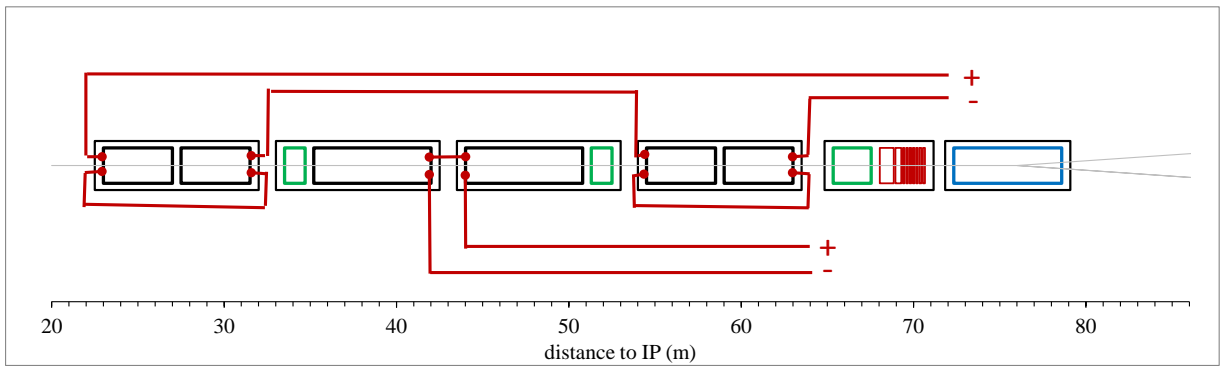

Figure 3-14: Sketch of baseline for triplet powering (trims not shown)

## 3.11 Cooling

The static heat load on the cold mass is mainly due to collision debris: 500 W on the triplet, and 70 W on the separation dipole. A factor of 1.5 has to be applied to these estimates to be compatible with the ultimate luminosity of $7.5 \times 10^{34}$ cm$^{-2}$ s$^{-1}$. The heat load on the triplet is removed via two 68 mm inner diameter heat exchangers, ultimately providing the ability to remove up to 1000 W (see Figure 3-15). To cope with these high heat loads, an additional low pressure pumping is added between Q2a and Q2b to keep the two-phase vapour flow velocity below 7 m/s, above which the HXs would not function correctly. These heat exchangers also cool the 1.2 m long orbit corrector. Simulations show that a solution with one (or more) heat exchangers cooling the whole string triplet–D1 is not viable. Therefore, a second system of heat exchangers is used to cool the corrector package and D1, which receive 50 W and 70 W, respectively at nominal luminosity. Here the baseline is to have two heat exchangers of 49 mm inner diameter, able to remove 250 W. One heat exchanger would provide only 125 W, barely enough to remove the 120 W due to collision debris, without the required 50% margin. Additional low pressure pumping is added between Q2a and Q2b.

The beam screen receives ~500 W in the triplet–correctors–D1 region (see Table 3-1, including 80 W from the interconnections). Given a 50% margin to reach ultimate peak luminosity, and a 150 W budget for the residual effect of electron cloud, the system has to remove ~975 W over 55 m, i.e. ~17 W/m. Heat is removed at 40–60 K [42]. The cooling tubes inner diameter is kept at ~7 mm due to an increase of the pressure of the helium to 18 bar. This choice is more challenging for the piping system but allows minimization of the space taken by the cooling pipes, which reduce the aperture available to the beam.

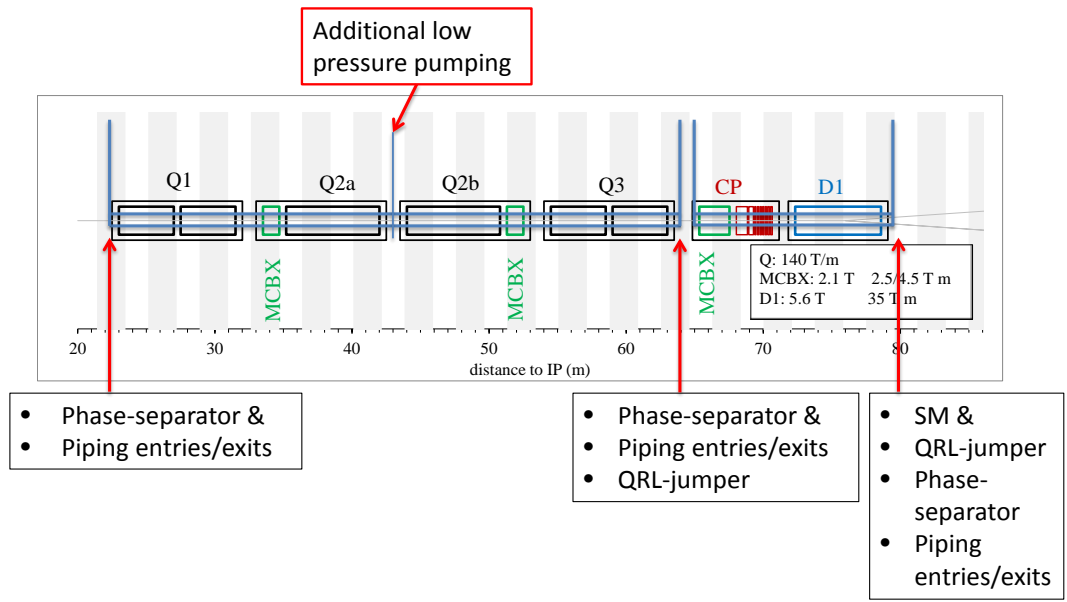

Figure 3-15: Sketch of the cooling system

For the stand-alone magnets, the jumper is located on the upper part of the slope, and the symmetry of the cryostat with respect to the interaction point is lost. This means that the left and right cryostats are not interchangeable, but the piping needed for the cooling system is simpler. This choice cannot be taken for the triplet, where there is no space for a jumper between the quadrupole and the interaction point. So the jumper is on the D1 side, requiring an additional return pipe.

## 3.12 Instrumentation

We give here a short summary of the instrumentation foreseen for the IR magnets.

- Quench protection requires voltage taps; for dipoles and quadrupoles we foresee the possibility of monitoring the voltage on each coil. Quench detection is therefore based on an analysis of the difference between signals of different coils, thus allowing cancelling of the inductive voltage.

- Beam loss monitors will be located inside the triplet cold mass to have the possibility of monitoring beam losses closer to the beam pipe and to the coil. Installation of special beam loss monitors in one of the iron holes not used by the heat exchangers or by the busbars is foreseen.

- Temperature sensors: one per cold mass, plus a spare, as in the LHC.

- Strain gauges: all quadrupoles based on the bladder and key structure will have strain gauges on the Al shell. Gauges will also be installed on the coil (winding pole) in the model and prototype phase, and possibly during production. One could consider having these signals extracted from the cryostat to have them available during operation. In the case of the LHC dipoles, strain gauges were used in the initial prototyping phase only.

- Beam position monitors will be placed in the interconnections between Q1 and Q2a, Q2b and Q3, Q3 and the corrector package, and between the corrector package and D1. Moreover they will be present at positions close to D2, Q4, and Q5.

## 3.13 Test

In general, magnets will be tested individually in a vertical test station, and then horizontally in the final cold mass assembly within the final cryostat. In some cases the first test will be possible in laboratories collaborating with CERN (for instance FNAL for Q1/Q3, and KEK for D1). The second test will be carried out at CERN. A string including the magnets from Q1 to D1 will be assembled in the CERN test facility (SM18) by 2019–2020. Magnetic measurements at 1.9 K with the rotating coil technique will be carried for all main magnets and for all the low-order (dipole and quadrupole) correctors. For the high-order corrector the magnetic measurement strategy is still to be defined.